\documentclass[12pt]{article}
\pdfoutput=1
\usepackage{geometry,enumerate,amsmath,amssymb}
\usepackage{fullpage}
\usepackage{graphicx}
\usepackage{bm}
\numberwithin{equation}{section}
\newcommand{\be}{\begin{equation}}
\newcommand{\ee}{\end{equation}}
\newcommand{\bea}{\begin{eqnarray}}
\newcommand{\eea}{\end{eqnarray}}
\renewcommand{\epsilon}{\varepsilon}

\begin{document}
\title{JNR Skyrmions}
\author{
  Paul Sutcliffe\\[10pt]
 {\em \normalsize Department of Mathematical Sciences,}\\
 {\em \normalsize Durham University, Durham DH1 3LE, United Kingdom.}\\ 
{\normalsize Email:  p.m.sutcliffe@durham.ac.uk}
}

\date{November 2024}

\maketitle
\begin{abstract}
  The JNR ansatz provides a simple formula to obtain families of charge $N$ self-dual $SU(2)$ Yang-Mills instantons in four-dimensional Euclidean space, from the free data of $N+1$ distinct points with associated positive weights. Here an analogous formula is presented for Skyrme fields in three-dimensional Euclidean space.  These families of Skyrme fields include good approximations to a range of Skyrmions, with energies that are typically within a few percent of the numerically computed solutions.
    \end{abstract}

\newpage

\section{Introduction}\quad
Skyrmions are topological solitons in three-dimensional Euclidean space, with a topological charge $N$ that is identified with baryon number \cite{Sk}. As there are no explicit exact solutions for Skyrmions, they must be studied using either numerical field theory computations or some other form of approximation for the Skyrme field. An approximation that is applicable for coincident Skyrmions is the rational map approximation \cite{HMS}. This yields a good description of the minimal energy Skyrmion of charge $N$, by specifying the angular dependence of the Skyrme field via a degree $N$ rational map between Riemann spheres. This is combined with a real radial profile function, $f(r)$, whose dependence on the distance $r$ from the centre of the Skyrmion can be calculated numerically by solving an ordinary differential equation. The rational map generated Skyrmion typically has an energy that is only  a few percent above that of the true minimal energy Skyrmion obtained from field theory computations.
The major disadvantage of the rational map approximation is that it cannot describe the deformation of the minimal energy charge $N$ Skyrmion into $N$ well-separated single Skyrmions. It is important to be able to model such a process in order to study both Skyrmion scattering and the quantization of Skyrmions beyond the rigid body approximation.

The Atiyah-Manton construction \cite{AM} approximates Skyrmions by the holonomy of $SU(2)$ Yang-Mills instantons in four-dimensional Euclidean space. The method to obtain a Skyrme field in three-dimensional space, with topological charge $N$, is to calculate the holonomy along lines parallel to the extra dimension of a charge $N$ instanton. The $8N$-dimensional moduli space of instantons generates a family of Skyrme fields that includes a good approximation to the minimal energy charge $N$ Skyrmion, together with parameters that allow this Skyrmion to be deformed into a collection of well separated single Skyrmions, and more complicated cluster decompositions.

The main drawback of the Atiyah-Manton approach is that the holonomy can only be calculated explicitly for the case of a spherically symmetric hedgehog Skyrme field, and in all other cases the holonomy must be computed numerically. Such numerical computations \cite{LM,CH} reveal that the typical error in the Atiyah-Manton approximation is less than two percent, when compared to field theory computations.
The reason that the Atiyah-Manton approximation works so well is that
Yang-Mills instanton holonomies produce exact solutions of an extended BPS Skyrme model, in which the Skyrme field is coupled to an infinite tower of vector mesons \cite{Su}. The Skyrme model is recovered from a truncation of this BPS Skyrme model by neglecting the vector mesons, explaining the accuracy of the Atiyah-Manton approximation in terms of the truncation of an exact correspondence.

The Yang-Mills instantons required as input for the Atiyah-Manton approximation can be obtained using the ADHM construction \cite{ADHM}. This provides a correspondence between the moduli space of instantons and the moduli space of certain quaternionic matrices, known as ADHM data, that satisfy a nonlinear algebraic constraint. Given ADHM data, the gauge potential of the Yang-Mills instanton can be calculated explicitly using only linear algebra.

Recently, a new approximation has been introduced \cite{HS} that provides explicit Skyrme fields directly from the ADHM data using only linear algebra, rather than computing the holonomy of the associated instanton. It is a Skyrmion analogue of the ADHM construction for instantons, although it provides approximations rather than exact solutions. The error is typically similar to that of the rational map approximation, so of the order of a few percent.
Skyrme fields obtained using this new approximation will be referred to as ADHM Skyrmions. The terminology ADHM Skyrmions \cite{CH} has previously been used differently to refer to Skyrme fields generated using the Atiyah-Manton approximation, because the instantons required as input to compute the holonomy in the Atiyah-Manton approximation can be specified by their ADHM data.
The two different meanings of the term ADHM Skyrmions are related, in that the new approximation was derived by performing an ultra-discretization of the line along which the instanton holonomy is calculated, replacing the interior points of the line by just three lattice points to turn a numerical scheme into a simple analytic formula \cite{HS}. The reason that such an extreme discretization works is that the ADHM construction involves an induced connection, and this implies a novel cancellation of the lattice spacing in the discretized version of the differential equation that defines the instanton holonomy.

In principle, all instantons can be obtained using the ADHM construction, but in practice the most general ADHM data cannot be obtained explicitly, because there is no general explicit solution for the nonlinear algebraic constraint for all $N$. A subset of charge $N$ instantons can be obtained using the JNR ansatz \cite{JNR}, which takes as input the free data of $N+1$ distinct points in $\mathbb{R}^4$, each with an associated positive real weight. This free data can be used to obtain corresponding explicit ADHM data, but the JNR ansatz provides a more direct route to obtain the Yang-Mills gauge potential, via the logarithmic derivative of a harmonic function with poles at the points specified by the free data. For $N=1$ and $N=2$ all instantons are of JNR type, but for $N>2$ they constitute a $(5N+7)$-dimensional submanifold of the $8N$-dimensional moduli space. The missing moduli, that grow in number like $3N$ for large $N$, may be viewed as independent $SU(2)$ orientations for each instanton.

The analogy with instantons suggests that from the ADHM construction of Skyrmions there should be a more direct formula for the Skyrme field for a class of JNR Skyrmions that are a subset of ADHM Skyrmions. The purpose of the present paper is to provide this formula for the Skyrme field in terms of the free data of poles and weights.

\section{JNR Skyrmions}\quad
A Skyrme field $U(x):\mathbb{R}^3\mapsto SU(2)$ is a smooth map that satisfies the boundary condition $U\to 1$ as $|x|\to\infty.$ It has an associated topological charge, $N\in\mathbb{Z}=\pi_3(SU(2))$, given by the formula
\be
N=\int \frac{1}{24\pi^2}\varepsilon_{ijk}\mbox{Tr}(R_iR_kR_j)\, d^3x,
\label{baryon}
\ee
where $R_i=\partial_iU\,U^{-1}$ are the $\mathfrak{su}(2)$-valued right currents for $i=1,2,3.$ 

In suitable units, the static Skyrme energy is 
\be
E=\frac{1}{12\pi^2}\int 
-\mbox{Tr}\bigg\{\frac{1}{2}R_i^2+\frac{1}{16}[R_i,R_j]^2
\bigg\}\,d^3x,
\label{energy}
\ee
and obeys the Faddeev-Bogomolny energy bound \cite{Fa}, which states that
$E\ge |N|$. This bound cannot be attained for $N\ne 0.$
The Skyrmions with minimal energies can be obtained from numerical field theory computations \cite{BS}, and the resulting energies are presented in Table 1 for the first few values of $N.$ For comparison, the energies of the approximations mentioned in the previous section are also provided.

\begin{table}[h!]
\begin{center}
  \begin{tabular}{|c|c|c|c|}
    \hline
    Method  & $N=1$ & $N=2$ & $N=3$ \\
    \hline
    Field theory & 1.232 & 2.358 & 3.438\\
    Atiyah-Manton & 1.243 & 2.384 & 3.488\\
    Rational maps & 1.232 & 2.416 & 3.552\\
    JNR & 1.236 &  2.418 & 3.554\\
    \hline
  \end{tabular}
\caption{The energies for minimal energy Skyrmions with charges $N=1,2,3$, obtained using numerical field theory computations and different approximate methods.}
\end{center}
\end{table}
In the following it will often be convenient to identify a point in $\mathbb{R}^3$ with a pure quaternion, $x=ix_1+jx_2+kx_3$,
a point in $\mathbb{R}^4$ with a quaternion, $X=x_4+x$,
and the $SU(2)$-valued Skyrme field with a unit quaternion.

The JNR ansatz \cite{JNR} for charge $N$ instantons gives a simple formula for the Yang-Mills gauge potential in terms of the harmonic function
\be
\Psi=\sum_{i=0}^N \frac{\lambda_i^2}{|X-a_i|^2}.
\label{JNRinstantons}
\ee
Here the $N+1$ poles, $a_i$, are distinct points in $\mathbb{R}^4$ (or equivalently quaternions) and $\lambda_i$ are the positive real weights. A noteworthy feature is that the number of poles is greater than the number of instantons.
In the limit $\lambda_0=|a_0|\to\infty$, the JNR form (\ref{JNRinstantons}) simplifies to
\be
\Psi=1+\sum_{i=1}^N \frac{\lambda_i^2}{|X-a_i|^2},
\label{tHinstantons}
\ee
which is known as the 't Hooft form, where the number of poles is now equal to the number of instantons. The interpretation of the $5N$ moduli in the 't Hooft form is that the poles give the positions of the instantons and the weights provide their sizes.

The main result in this paper is the following formula for a charge $N$ Skyrme field, given the same JNR data that appears in (\ref{JNRinstantons}),
\be
U=\frac{\psi(-\mu)\overline{\psi(\mu)}}{|\psi(-\mu){\psi(\mu)}|},
\label{psi2u}
\ee
where
 \be  
 \psi(\mu)=  \bigg|\sum_{i=0}^N \frac{\lambda_i^2(x-a_i)}{|x-a_i|^2}\bigg|^2\sum_{j=0}^N \frac{\lambda_j^2}{|x-a_j+\mu|^2}
 + \sum_{i,j,k=0}^N\frac{\mu\lambda_i^2\lambda_j^2\lambda_k^2(a_i-a_j)(x-a_i)(\bar a_i-\bar a_k)}{|x-a_i+\mu|^2|x-a_j+\mu|^2|x-a_i|^2|x-a_k|^2}.
 \label{sympsi}
  \ee
  In the above $\mu$ is a positive real parameter that can be optimized to minimize the energy of the Skyrme field for fixed JNR data. Note that $\psi$ is a quaternionic-valued function on $\mathbb{R}^3$, with argument $x$ that is suppressed for clarity, whereas $\Psi$ is a real-valued function on $\mathbb{R}^4$, with argument $X$. However, for a given $\mu$, both functions contain the same parameters and have similar pole structures.
  
  The formula (\ref{sympsi}) for JNR Skyrmions will be derived in Section 3, from the ADHM construction of Skyrme fields introduced in \cite{HS}. As all ADHM data is equivalent to JNR data for $N=1$ and $N=2$, all the examples presented in \cite{HS} can be reproduced using (\ref{sympsi}). The details will be provided for some illustrative cases. The advantage of the JNR approach over the ADHM construction is that it is provides a direct formula for the Skyrme field that applies to free data and avoids the requirement to invert an $N\times N$ matrix. The disadvantage is that it provides a smaller family of Skyrme fields for $N>2.$ The first significant Skyrme field that is inside the ADHM family but lies outside the JNR family is an approximation to the minimal energy $N=4$ Skyrmion with cubic symmetry.

  Note that there is an equivalence relation $\sim$ on $\psi(\mu)$, because it may be multiplied by any real quantity without changing the associated Skyrme field $U$ obtained from (\ref{psi2u}). Using this equivalence the 't Hooft form can be obtained by taking the limit $a_0=\lambda_0\to \infty$, giving
  \be
  \psi(\mu)\sim \psi(\mu)/\lambda_0^2\to
  1+\sum_{i=1}^N \frac{\lambda_i^2(x-a_i)(\bar x-\bar a_i +\mu)}{|x-a_i|^2|x-a_i+\mu|^2}.
  \label{tH}
  \ee
  More generally, if the limit is taken as $|a_0|=\lambda_0\to \infty$, then the same Skyrme field is obtained but with an isospin rotation, corresponding to replacing $U$ with $q\,U\bar q$, where $q$ is the unit quaternion $q=a_0/|a_0|.$
  
  The analysis of the 't Hooft form is simplified by taking all the poles to be pure quaternions, $\bar a_i=-a_i$ for $i=1,...,N.$ This is a rather natural restriction for Skyrmions in $\mathbb{R}^3$, as the poles can then be identified with the positions of the Skyrmions, in a similar way to the identification of instanton positions in $\mathbb{R}^4$. In the case of pure quaternion poles the 't Hooft form has the obvious symmetry $\overline{\psi(\mu)}=\psi(-\mu)$,
because $|x-a_i+\mu|^2=|x-a_i|^2+\mu^2,$  
  and therefore the expression (\ref{psi2u}) simplifies to
\be
U=\bigg(\frac{\psi(-\mu)}{|\psi(-\mu)|}\bigg)^2.
\label{psi2u2}
\ee
This formula provides a Skyrme field that is a rational function of the Cartesian coordinates, whereas (\ref{psi2u}) only implies an algebraic function in general.

Taking $\psi(\mu)$ given by the 't Hooft form (\ref{tH}), it is clear that $\psi(\mu)\to 1$ as $|x|\to\infty$, hence the required boundary condition that $U\to 1$ is satisfied. By definition, the positions of the Skyrmions are the points in space where $U=-1.$ For pure quaternion poles (\ref{psi2u2}) shows that $U=-1$ corresponds to $\psi(-\mu)$ being a pure quaternion. As the real part of (\ref{tH}) is positive these positions are independent of $\mu$ and correspond to the locations of the poles $a_i$ of $\psi(\mu)$, because $\psi(\mu)/|\psi(\mu)|$ tends to a pure quaternion as any of the poles are approached. This agrees with the fact that the topological charge $N$ can also be obtained by counting the number of preimages of $U=-1.$

Returning now to the JNR case, the restriction to pure quaternion poles $\bar a_i=-a_i$ for all $i=0,...,N,$ is again natural, and is often the situation of most interest for obtaining low energy Skyrme fields.
The following analysis applies to this restricted case of pure quaternion poles.
Introducing a notation for each of the two terms in (\ref{sympsi}), by writing $\psi(\mu)={\cal R}(\mu)+{\cal I}(\mu)$, it is clear that the real function
\be
   {\cal R}(\mu)=\bigg|\sum_{i=0}^N \frac{\lambda_i^2(x-a_i)}{|x-a_i|^2}\bigg|^2\sum_{j=0}^N \frac{\lambda_j^2}{|x-a_j+\mu|^2},
   \label{realpart}
   \ee
   is now an even function of $\mu$.
   It is also clear that the quaternionic-valued function
   \be
    {\cal I}(\mu)=
   \sum_{i,j,k=0}^N\frac{\mu\lambda_i^2\lambda_j^2\lambda_k^2(a_i-a_j)(x-a_i)(\bar a_i-\bar a_k)}{|x-a_i+\mu|^2|x-a_j+\mu|^2|x-a_i|^2|x-a_k|^2},
   \label{qpart}
   \ee
   is an odd function of $\mu$.
   To show that ${\cal I}(\mu)$ is a pure quaternion it is convenient to introduce the notation $y_i=x-a_i=-\bar y_i$. The real part of ${\cal I}(\mu)$ may be written as
   \bea
      \mbox{Re\ }{\cal I}(\mu)
      &=& \mbox{Re} \sum_{i,j,k=0}^N
      \frac{\mu\lambda_i^2\lambda_j^2\lambda_k^2(y_i-y_j)y_i(\bar y_i-\bar y_k)}{|y_i+\mu|^2|y_i|^2|y_j+\mu|^2|y_k|^2}    
      \\
                  &=& \mbox{Re} \sum_{i,j,k=0}^N
      \frac{\mu\lambda_i^2\lambda_j^2\lambda_k^2y_jy_i\bar y_k}{|y_i+\mu|^2|y_i|^2|y_j+\mu|^2|y_k|^2}    \\
      &=&
      \mbox{Re} \sum_{i,j,k=0}^N\frac{\mu\lambda_i^2\lambda_j^2\lambda_k^2(|y_k|^2+\mu^2)|y_j|^2{y_jy_i\bar y_k}}
           {|y_i+\mu|^2|y_i|^2|y_j+\mu|^2|y_j|^2|y_k+\mu|^2|y_k|^2} \label{pen}\\
      &=&
      \mbox{Re} \sum_{i,j,k=0}^N\frac{\mu^3\lambda_i^2\lambda_j^2\lambda_k^2|y_j|^2{y_jy_i\bar y_k}}
           {|y_i+\mu|^2|y_i|^2|y_j+\mu|^2|y_j|^2|y_k+\mu|^2|y_k|^2} =0,   \label{last}  
      \eea
      where (\ref{last}) is obtained from (\ref{pen}) after applying a relabelling that swaps $j$ and $k$ in the term proportional to
      $|y_k|^2|y_j|^2y_jy_i\bar y_k = -|y_k|^2|y_j|^2\overline{y_ky_i\bar y_j}$.
The final expression (\ref{last}) vanishes after applying a relabelling that swaps $i$ and $k$, because $y_i\bar y_k+y_k\bar y_i$ is real.
      
Taken together, the results that ${\cal R}(\mu)={\cal R}(-\mu)=\overline{{\cal R}(\mu)}$ and ${\cal I}(\mu)=-{\cal I}(-\mu)=-\overline{{\cal I}(\mu)}$, shows that the symmetry property, $\overline{\psi(\mu)}=\psi(-\mu)$, is also satisfied in the JNR case, for pure quaternion poles. Therefore the simplified formula
(\ref{psi2u2}) producing a rational Skyrme field again applies.
However, the poles $a_i$ of (\ref{sympsi}) no longer correspond to the positions of the Skyrmions, because both ${\cal R}(\mu)$ and ${\cal I}(\mu)$ have poles at $a_i$, so $\psi(\mu)/|\psi(\mu)|$ does not tend to a pure quaternion as any of the poles are approached. Of course, this must be the situation because there are $N+1$ poles and the number of preimages of $U=-1$ cannot exceed $N$. The Skyrmion positions are given by the vanishing of ${\cal R}(-\mu)$. These positions are independent of $\mu$ and are given by the zeros of $\zeta$, defined to be
\be
\zeta = \sum_{i=0}^N \frac{\lambda_i^2(x-a_i)}{|x-a_i|^2}.
\label{zeta}
\ee
In the regime that one weight, say $\lambda_j$, is much larger than all the others, $\lambda_j\gg \lambda_i$, for all $i\ne j$, formula (\ref{zeta}) shows that the Skyrmion positions are close to the $N$ poles $a_i$ with $i\ne j$, associated with the other weights.

To obtain an $N=1$ Skyrmion as a JNR Skyrmion, set $\lambda_0=\lambda_1=1$ and $a_0=-a_1=q$, to fix the position at the origin, with $q$ a pure quaternion. This gives
\be
\zeta=\frac{(x-q)\bar x(x+q)}{|x-q|^2|x+q|^2},
\ee
which indeed vanishes if and only if $x=0.$ Using the JNR formula (\ref{sympsi})
yields
\be
\psi(\mu)\sim |x|^2(|x|^2+|q|^2+\mu^2)+\mu qx\bar q.
\label{jnrpsi1}
\ee
Setting $|q|=\mu^2=2$, the Skyrme field obtained by substituting (\ref{jnrpsi1}) into
(\ref{psi2u2}) is
\be
U=
\frac{r^{6}+12 r^{4}+36 r^{2}-32-2\sqrt{2}(r^{2}+6)q x\bar q}{\left(r^{2}+8\right) \left(r^{2}+2\right)^{2}},
\ee
where $r=|x|$ is the distance from the origin.
This reproduces the hedgehog Skyrme field from the ADHM construction \cite{HS}, with the energy given in Table 1, and an isospin orientation determined by the unit quaternion $q/|q|,$ that specifies the direction of the line through the two poles. This formula is also valid for real $q$, so setting $q=2$ gives a hedgehog field in standard orientation.

The ADHM generated $N=2$ axially symmetric Skyrme field \cite{HS}, with the energy given in Table 1, is reproduced in JNR form by taking $\mu=2$ and unit weights with poles on the vertices of an equilateral triangle
\be
a_0=2i, \ \ a_1=-i+\sqrt{3}j, \ \ a_2=-i-\sqrt{3}j.
\ee
There is axial symmetry, rather than just triangular symmetry, because rotating the triangle simply produces an isospin rotation. $\zeta$ vanishes at the origin, the position of the Skyrmion, because the poles have equal magnitude and their sum vanishes.

The minimal energy $N=3$ Skyrmion has tetrahedral symmetry. It is approximated by a JNR Skyrmion with $\mu=2$ and unit weights with poles on the vertices of a regular tetrahedron
\be
a_0=\frac{5}{4}(i+j+k), \ \
a_1=\frac{5}{4}(-i-j+k), \ \
a_2=\frac{5}{4}(-i+j-k), \ \
a_3=\frac{5}{4}(i-j-k).
\label{tet3}
\ee
The energy is presented in Table 1 and is again very close to that of the rational map approximation.

Having discussed several illustrative examples of JNR Skyrmions, it is now time to turn to the derivation of the main formula (\ref{sympsi}).

\section{JNR from ADHM}\quad
ADHM data consists of a matrix
\be
\widehat M = \begin{pmatrix} L\\ M \end{pmatrix},
\ee
where $L$ is a row of $N$ quaternions and $M$ is a symmetric $N\times N$ matrix of quaternions. ADHM data must satisfy the condition that
the $N\times N$ matrix $\widehat M^\dagger \widehat M$
is real and non-singular, where ${}^\dagger$ denotes the quaternionic conjugate transpose.

For notational convenience, in the following the full dependence on $X=x+x_4$ will be suppressed, with only the $x_4$  dependence written explicitly as a functional argument, with the $x$ dependence left implicit. The ADHM operator is 
\be
\Delta(x_4)=\begin{pmatrix}L\\ M-1_{N}X\end{pmatrix},
\label{adhmoperator}
\ee
where $1_N$ denotes the $N\times N$ identity matrix. A requirement of the ADHM construction is that $\Delta(x_4)^\dagger\Delta(x_4)$ is a real non-singular matrix.
Defining the $(N+1)\times (N+1)$ quaternionic projector matrix
\be
Q(x_4)=1_{N+1} -\Delta(x_4)\bigg(\Delta(x_4)^\dagger\Delta(x_4)\bigg)^{-1}\Delta(x_4)^\dagger,
\label{defQ}
\ee
the ADHM Skyrme field is given by \cite{HS}
\be
U
=\frac{e_1^t Q(-\mu)Q(0)Q(\mu)e_1}
{|e_1^t Q(-\mu)Q(0)Q(\mu)e_1|},
\label{scheme5}
\ee
where $e_1$ is the $(N+1)$-component column vector with first entry equal to 1 and all other entries equal to $0$.
As in the previous section, $\mu$ is a positive real parameter that can be optimized to minimize the energy of the Skyrme field.
This parameter appears in the derivation of (\ref{scheme5}) via the discretization of the line parameterized by $x_4$, where the interior points of this line are replaced by the three points $x_4=\pm\mu,0$ in a lattice formulation of the instanton holonomy.

To restrict to JNR data, the operator $\Delta(x_4)$ is taken to be
\be
\Delta(x_4)=S\,\Gamma(x_4)V,
\label{changeb}
\ee
with
\be
\Gamma(x_4)=
\begin{pmatrix}
\lambda_1 (a_0-X) & \lambda_2 (a_0-X) & \cdots & \lambda_N (a_0-X)\\
\lambda_0 (a_1-X) & & & \\
& \lambda_0 (a_2-X) & & \\
& & \ddots & \\
& & & \lambda_0 (a_N-X)
\end{pmatrix},
\label{defGamma}
\ee
where $S\in O(N+1)$ and $V\in GL(N,\mathbb{R})$
are matrices that satisfy
\be 
S
\begin{pmatrix}
\lambda_1  &  \lambda_2 & \cdots & \lambda_N \\
\lambda_0  & & & \\
& \lambda_0  & & \\
& & \ddots & \\
& & & \lambda_0 
\end{pmatrix}
V
= \begin{pmatrix}
0  &  0 & \cdots & 0 \\
1  & & & \\
& 1  & & \\
& & \ddots & \\
& & & 1
\end{pmatrix},
\ee
with explicit formulae for their entries given in \cite{BCS}.
Labelling the entries of $S$ as $S_{ij}$ with $i,j=0,...,N$, the only values that will be required in the following are
\be
S_{0i}=\frac{\lambda_i\sigma_i}{\sqrt{\sum_{j=0}^N \lambda_j^2}},
\qquad \mbox{ where } \quad
\sigma_i=\begin{cases} \ \ 1 & \mbox{ if } i=0 \\ -1 & \mbox{ if } i\ne 0.
\end{cases}
\label{S0i}
\ee

Substituting (\ref{changeb}) into (\ref{defQ}), and the result into (\ref{scheme5}) gives
\be
U
=\frac{e_1^t S P(-\mu)P(0)P(\mu)S^te_1}
{|e_1^t S P(-\mu)P(0)P(\mu)S^te_1|},
\label{nscheme5}
\ee
where
\be
P(x_4)=1_{N+1} -\Gamma(x_4)\bigg(\Gamma(x_4)^\dagger\Gamma(x_4)\bigg)^{-1}\Gamma(x_4)^\dagger.
\label{defP}
\ee
The simple form of (\ref{defGamma}) allows the entries of $P$ to be calculated to be
\be
P_{ij}(x_4)=\frac{\lambda_i\lambda_j\sigma_i\sigma_j(X-a_i)\overline{(X-a_j)}}
{|X-a_i|^2|X-a_j|^2\sum_{k=0}^N\lambda_k^2/|X-a_k|^2}
\sim
\frac{\lambda_i\lambda_j\sigma_i\sigma_j(X-a_i)\overline{(X-a_j)}}
     {|X-a_i|^2|X-a_j|^2},
     \label{Pij}
\ee
where $\sim$ exploits the equivalence relation that $P$ may be multiplied by any positive real quantity without changing the associated Skyrme field $U$ obtained from (\ref{nscheme5}).

Using (\ref{Pij}) the numerator in (\ref{nscheme5}) becomes
\begin{multline}
e_1^t S P(-\mu)P(0)P(\mu)S^te_1 =\\
\sum_{i,j,k,l=0}^N
\frac{
\lambda_i\lambda_\ell\lambda_j^2\lambda_k^2\sigma_i\sigma_l S_{0i}S_{0\ell}
  (x-a_i-\mu)\overline{(x-a_j-\mu)}
  (x-a_j)\overline{(x-a_k)}
  (x-a_k+\mu)\overline{(x-a_\ell+\mu)}
}
{
  |x-a_i-\mu|^2|x-a_j-\mu|^2
  |x-a_j|^2|x-a_k|^2
  |x-a_k+\mu|^2|x-a_\ell+\mu|^2
}
\\
\sim
\sum_{i,j,k,l=0}^N
\frac{
\lambda_i^2\lambda_j^2\lambda_k^2\lambda_\ell^2
  (x-a_i-\mu)\overline{(x-a_j-\mu)}
  (x-a_j)\overline{(x-a_k)}
  (x-a_k+\mu)\overline{(x-a_\ell+\mu)}
}
{
  |x-a_i-\mu|^2|x-a_j-\mu|^2
  |x-a_j|^2|x-a_k|^2
  |x-a_k+\mu|^2|x-a_\ell+\mu|^2
},
\end{multline}
where the final line follows after using (\ref{S0i}). Factorizing this expression yields (\ref{psi2u}) with
\be
\psi(\mu)=\sum_{i,j=0}^N
\frac{
\lambda_i^2\lambda_j^2
  (x-a_i+\mu)\overline{(x-a_j+\mu)}(x-a_j)
}
{
  |x-a_i+\mu|^2|x-a_j+\mu|^2
  |x-a_j|^2
}.
\ee
This is not the expression given in (\ref{sympsi}), and in particular it does not have the reality properties exploited in Section 2. However, note that the formula (\ref{psi2u}) relating $\psi(\mu)$ and $U$ not only allows for the equivalence relation that $\psi(\mu)$ may be multiplied by any real quantity, but also allows $\psi(\mu)\sim \psi(\mu) \chi$, where $\chi$ is any quaternionic quantity that is independent of $\mu.$
Exploiting this equivalence with
\be
\chi=\sum_{k=0}^N \frac{\lambda_k^2\overline{(x-a_k)}}{|x-a_k|^2},
\ee
gives
\bea
\psi(\mu)&=&
\sum_{i,j,k=0}^N
\frac{
\lambda_i^2\lambda_j^2\lambda_k^2
  (x-a_j+\mu+a_j-a_i)\overline{(x-a_j+\mu)}(x-a_j)\overline{(x-a_k)}
}
{
  |x-a_i+\mu|^2|x-a_j+\mu|^2
  |x-a_j|^2|x-a_k|^2
}\nonumber\\
&=&
\sum_{i,j,k=0}^N
\frac{
\lambda_i^2\lambda_j^2\lambda_k^2
  (x-a_j)\overline{(x-a_k)}
}
{
  |x-a_i+\mu|^2
  |x-a_j|^2|x-a_k|^2
}
+\frac{
\lambda_i^2\lambda_j^2\lambda_k^2
  (a_j-a_i)\overline{(x-a_j+\mu)}(x-a_j)\overline{(x-a_k)}
}
{
  |x-a_i+\mu|^2|x-a_j+\mu|^2
  |x-a_j|^2|x-a_k|^2}
\nonumber\\
&=&
{\cal R}(\mu)
+{\cal J}(\mu),
\eea
where the first term has been recognized as the real term ${\cal R}(\mu)$ defined in (\ref{realpart}), and the second term has been defined to be
\be
   {\cal J}(\mu)=
\sum_{i,j,k=0}^N
   \frac{
\lambda_i^2\lambda_j^2\lambda_k^2
  (a_j-a_i)\overline{(x-a_j+\mu)}(x-a_j)\overline{(x-a_k)}
}
{
  |x-a_i+\mu|^2|x-a_j+\mu|^2
  |x-a_j|^2|x-a_k|^2}.
\label{defJ}
\ee
   It now remains to show that ${\cal J}(\mu)$ is equal to ${\cal I}(\mu)$, defined in (\ref{qpart}). The linear dependence on $\mu$ of the numerator in (\ref{defJ}) prompts the splitting of ${\cal J}(\mu)$ into two terms
\be
    {\cal J}(\mu)=
    \sum_{i,j,k=0}^N
    \frac{\lambda_i^2\lambda_j^2\lambda_k^2(a_j-a_i)\overline{(x-a_k)}}
{|x-a_i+\mu|^2|x-a_j+\mu|^2|x-a_k|^2}
+
\frac{\mu\lambda_i^2\lambda_j^2\lambda_k^2(a_j-a_i)(x-a_j)\overline{(x-a_j+a_j-a_k)}}
{|x-a_i+\mu|^2|x-a_j+\mu|^2|x-a_j|^2|x-a_k|^2}.
\ee
The first term vanishes because the summand is antisymmetric under the exchange of $i$ and $j$, therefore
\be
   {\cal J}(\mu)=
   \sum_{i,j,k=0}^N
   \frac{\mu\lambda_i^2\lambda_j^2\lambda_k^2(a_j-a_i)}
        {|x-a_i+\mu|^2|x-a_j+\mu|^2|x-a_k|^2}
        +
        \frac{\mu\lambda_i^2\lambda_j^2\lambda_k^2(a_j-a_i)(x-a_j)(\bar a_j-\bar a_k)}
             {|x-a_i+\mu|^2|x-a_j+\mu|^2|x-a_j|^2|x-a_k|^2}.
             \ee
Again the first term vanishes because the summand is antisymmetric under the exchange of $i$ and $j$. The remaining term is recognized as the expression (\ref{qpart}) for ${\cal I}(\mu)$, after a relabelling of the indices that swaps $i$ and $j$. This completes the derivation of the JNR Skyrmion formula (\ref{sympsi}).          

\section{Outlook}\quad
JNR Skyrmions, and more generally ADHM Skyrmions, have an algebraic decay that is appropriate for the Skyrme model with massless pions. However, Skyrmions in models with massive pions decay exponentially. It would be interesting to find a modification of either the JNR or ADHM construction that is appropriate for massive pions. The structure of the 't Hooft form (\ref{tH}), as a sum of poles giving the positions of the Skyrmions, suggests that a first step is to write the hedgehog Skyrmion,
\be
U=\cos f(r)-\hat x\sin f(r),
\label{hh}
\ee
in a similar form. Here $\hat x=x/|x|$ and $f(r)$ is the radial profile function of the Skyrmion, a function of $r=|x|$, which could be the numerically calculated function in a model of either massless or massive pions. It is easy to check that (\ref{hh}) can be rewritten as $U=\phi^2/|\phi|^2$, where
\be
\phi=1-\frac{\hat x\sin f(r)}{1+\cos f(r)}
\label{sup1}
\ee
has similarities to the $N=1$ 't Hooft form (\ref{tH}). Note that $f(r)\to 0$ as $r\to\infty$, so $\phi\to 1$ and the boundary condition $U\to 1$ is satisfied.
Furthermore, $f(r)\to \pi$ as $r\to 0$, so $\phi$ has a pole at the origin and $\phi/|\phi|$ tends to a pure quaternion as the pole is approached, giving $U=-1$ at $r=0$. This way of writing the hedgehog Skyrmion therefore shares the qualitative features of the $N=1$ 't Hooft Skyrmion but allows the decay to be controlled by specifying a profile function.

There is an obvious generalization of (\ref{sup1}) to $N>1$ by adding more poles. Using the same notation as earlier, $y_i=x_i-a_i$, with pure quaternion poles,
set
\be
\phi=1-\sum_{i=1}^N\frac{q_i\hat y_i \bar q_i\sin f(|y_i|)}{1+\cos f(|y_i|)},
\label{sup}
\ee
where the unit quaternions $q_i$ have been introduced to allow independent isospin rotations for each of the Skyrmions. This is a new way to obtain a Skyrme field for a superposition of well-separated Skyrmions.

The standard method to obtain the Skyrme field of a pair of well-separated Skyrmions is to use the product
ansatz $U=U_1U_2$, where $U_1$ and $U_2$ are the hedgehog fields of single Skyrmions with well-separated positions. The problem with this ansatz is that $U_1U_2\neq U_2U_1$, so it breaks the symmetry of a true superposition.
The symmetrized product ansatz \cite{NR} was introduced to restore this symmetry. It is given by $U=(U_1U_2+U_2U_1)/|U_1U_2+U_2U_1|$, but only produces a smooth charge two Skyrme field for sufficiently large separations, because the denominator can vanish otherwise.

The superposition formula (\ref{sup}) is a new way to preserve the symmetry, as the fundamental operation is the addition of quaternions, rather than multiplication. It generates a smooth charge $N$ Skyrme field for any set of distinct poles. However, in common with both the product ansatz and its symmetrized version, the superposition only provides a good approximation to a low energy Skyrme field for well-separated Skyrmions. Numerical investigations of a pair of Skyrmions, in an isospin orientation that produces the most attractive channel, reveal that the energy as a function of separation using (\ref{sup}) lies between that of the product ansatz and the symmetrized product ansatz. The energy is close to that of the symmetrized product ansatz, and is therefore a reasonable alternative, with a slight advantage that the positions of the Skyrmions are given precisely by the poles, whereas the Skyrmion positions from the symmetrized product ansatz are only close to the positions of the individual Skyrmions used as input.

The superposition formula (\ref{sup}) was motivated by the 't Hooft form (\ref{tH}), which does not provide a good description of low energy Skyrmions as they merge. It is therefore not surprising that it is only applicable for well-separated Skyrmions. It would be useful to find a similar superposition formula based on the JNR form (\ref{sympsi}), which should then be able to describe Skyrmions for all separations in any Skyrme model, including massive pions, given the relevant profile function. This is currently under investigation.

In the quantization of Skyrmions the Finkelstein-Rubinstein constraints \cite{FR} play a vital role. These state that under a closed loop in the configuration space of charge $N$ Skyrme fields the wavefunction should acquire a minus sign if and only if the loop is non-contractible. Recently, elegant formulae have been derived \cite{CoHa} to determine whether loops described by ADHM data or rational maps are contractible. For JNR Skyrmions with poles of equal weight, a simple way to generate a closed loop is to permute the poles. It can be shown that this loop is contractible if and only if the permutation is even \cite{Cork}. The simplest example is for the $N=1$ JNR Skyrmion described earlier, where a path that rotates the poles $a_0$ and $a_1$ around a circle through an angle $\pi$ results in the odd permutation that swaps $a_0$ and $a_1$. This is the non-contractible loop that rotates the Skyrmion through an angle $2\pi$ and the change of sign of the wavefunction quantizes the Skyrmion as a fermion.

Taking the $N=2$ 't Hooft form with equal weights $\lambda_1=\lambda_2=2$ and the two poles $a_1=-a_2=q$, where $q$ is a pure quaternion with $|q|\gg 1$, gives a pair of well-separated charge one Skyrmions. The path that rotates the poles $a_1$ and $a_2$ around a circle through an angle $\pi$ yields the odd permutation that swaps $a_1$ and $a_2$. This is the non-contractible loop that exchanges the pair of Skyrmions, in agreement with the spin–statistics theorem.

An example of a more complicated non-contractible loop discussed in \cite{CoHa}, in terms of both ADHM data and rational maps, is the deformation of the tetrahedral $N=3$ Skyrmion via an axially symmetric charge three Skyrmion. As a JNR Skyrmion with equal weights this can be described by the path $t\in[0,1]$ with
\bea
a_0&=&\frac{5}{4}((i+j)\cos(\pi t/2)+(i-j)\sin(\pi t/2)+k(1-2t)), \nonumber\\
a_1&=&\frac{5}{4}(-(i+j)\cos(\pi t/2)-(i-j)\sin(\pi t/2)+k(1-2t)),\nonumber\\
a_2&=&\frac{5}{4}(-(i-j)\cos(\pi t/2)+(i+j)\sin(\pi t/2)-k(1-2t)), \nonumber\\
a_3&=&\frac{5}{4}((i-j)\cos(\pi t/2)-(i+j)\sin(\pi t/2)-k(1-2t)).
\eea
The path begins at $t=0$ as the tetrahedral JNR data (\ref{tet3}). At the midpoint of the path, $t=1/2$, the four poles are on the vertices of a square, giving an axially symmetric Skyrme field. The JNR data at the end of the path, $t=1$, is obtained from the JNR data at the start of the path by applying the odd permutation
$3201=(02)(01)(03)$, confirming that the loop is non-contractible. 

\section{Conclusion}\quad
By making use of a recently introduced ADHM construction for Skyrme fields \cite{HS}, a direct formula for JNR Skyrme fields has been derived. The formula simplifies if all the JNR poles are taken to lie in $\mathbb{R}^3$ and produces Skyrme fields that are rational functions of the Cartesian coordinates. The positions of the Skyrmions are given by the vanishing of a simple sum of poles.
JNR Skyrmions provide good approximations to a range of low charge Skyrmions, with errors in the energies of the order of a few percent, which is comparable to the rational map approximation.

Motivated by the qualitative features of the JNR Skyrmion formula in the 't Hooft limit, a new superposition formula has been presented that is applicable to well-separated Skyrmions and respects the symmetry that is broken by the standard product ansatz. There is hope that future work may provide a similar formula that is inspired by the general JNR case, rather than the 't Hooft limit, so that the well-separated restriction can be removed. This would allow the method to be applied to variants of the Skyrme model, including those with massive pions.

\end{document}